# Simulation of the transient photocurrent response of polycrystalline photocells based on equivalent circuit analysis

B. Vainas[*]

Department of Materials and Interfaces, the Weizmann Institute of Science, Rehovot 76100, Israel

**Abstract:**

We propose an equivalent circuit representation of the photogenerated charge separation and propagation in dye sensitized polycrystalline semiconductor in contact with a redox electrolyte. The suggested equivalent circuit for this type of photocell is based on an electrical transmission-line, and uses distributed photodiode model for the semiconductor-redox electrolyte interface. It was found that for small signal conditions, diodes can be replaced by equivalent elements, each consisting of a resistor and a capacitor connected in series. The equivalent circuit also provides for Beer-Lambert characteristics of light absorption.
The simulation of the transmission line equivalent circuit, subjected to short pulse illumination, allows us to reproduce the experimentally found difference between the photocurrent responses of the photocell to illumination from the electrolyte side, and its response to illumination through the semitransparent back electrode, to which the polycrystalline semiconductor layer is attached.
We suggest an electrochemical model for photogenerated charge separation, whereby the electrical field across the Helmholtz electrostatic double layer at the polycrystalline phase – electrolyte interface separates the photogenerated carriers.

**Introduction:**

An obvious objection to the field dependent photogenerated charge separation model at the polycrystalline phase surface – electrolyte interface is based on the argument, that for large area capacitive interfaces that are electrically in series with smaller area capacitors at the back interface, an applied voltage bias falls across the smaller capacitor. Another argument, specific to nanoscale polycrystalline layers, says that size characteristics of nanocrystalline semiconductors, prevents the built-up of a space charge, hence no significant electric fields can be supported.
Here we study a model based on an ohmic back contact. Consider the case where the interface area between the polycrystalline material and the electrolyte can be much larger than the contact area between the polycrystalline layer and the semitransparent back contact. For a non-capacitive, ohmic back contact, the very large interface capacitor between the polycrystalline material and the electrolyte, which is in series with the ohmic back contact, can be charged upon an application of an external voltage bias. Similar charging occurs in double-layer (ultra)capacitors where one can get about ~1V voltage across the electrochemical double layer. Typical values of capacitance of ultracapacitors, (or supercapacitors) are in the order of 1 F for 1 $cm^3$ of cell volume.

---

[*] On leave from Soreq Nuclear Research Center, Yavne 81800, Israel.

As to the argument concerning space charge formation in nanocrystals, we suggest that the electrochemical, double-layer charge distribution provides the electric field for photogenerated charge separation. The charge on the nanocrystalline side of the double layer can be accommodated in surface states, while the countercharges are ions in the electrolyte. As will be shown below, the small signal response of the diode-based transmission line suggests the participation of trap-like circuit elements, which can rationalize the surface states model for the semiconductor's side of the Helmholtz double layer.

The dye molecules, attached to the surface of the polycrystalline material (the nanonocrystalline $TiO_2$ in the particular system simulated here) are then subjected to the electrical field of the Helmholtz double layer. If this qualitative picture is correct, we can justify electrical field separation of photogenerated charges at the nanocrystalline – electrolyte interface, even if no space charge can be accommodated in the bulk of the nanocrystalline semiconductor.

We have constructed a SPICE-type schematics [1] for the transient response of the dye sensitized nanocrystalline $TiO_2$ electrolytic cell. The equivalent circuit is simply a transmission line of connected photodiode elements, each characterized by a combination of a current source followed by a resistor, in series, and a combination of a diode and a capacitor in parallel to it [2]. The small load resistor, the voltage drop on which is proportional to the photocurrent, sets the short circuited conditions. The small load resistor is connected directly to the end of the transmission line, hence an effective ohmic back contact.

**Results:**

The simulation of the transient photoresponse was done on the equivalent circuit given in Fig. 1 below, which represents a sequence of ten photodiode elements. It represents the case of illumination from the electrolyte side of the photocell, which is expressed through the constant fraction (0.7 in our simulations) reduction of currents supplied by the current sources. This current distribution simulates the exponential reduction of light intensity as light is absorbed while propagating through the material, according to the Beer – Lambert law. Note that the resistive load (R15) in Fig. 1 is attached to the rightmost current source of the transmission line, simulating the current collection from the farthest, in terms of the illumination source, light absorbing layer of the photocell. We call this a "front" (F) setup. Given the small value of the measuring resistor R15, the simulation of the equivalent circuit in Fig. 1 is done under short circuit conditions.

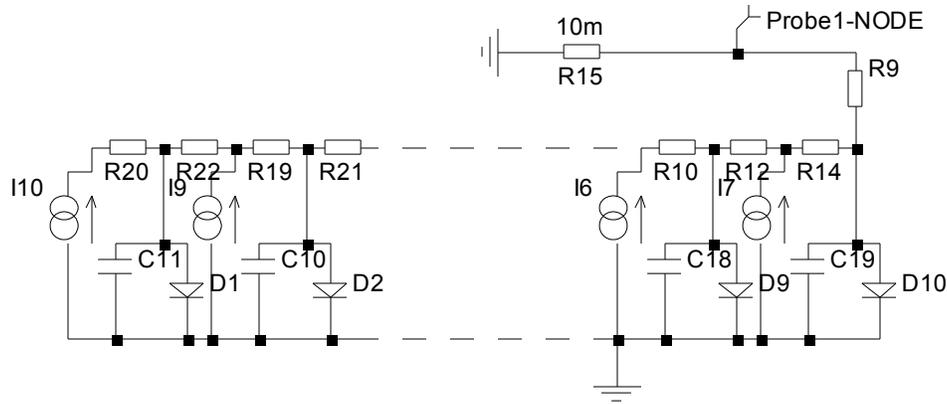

Fig. 1: The SPICE schematics of the dye sensitized nanocrystalline $TiO_2$ electrolytic cell. All diodes are DIN4148, all resistors apart from the load resistor, R15 (10 mohm), are 1 ohm, and all capacitors are 1.8 pF. The pulse duration (time interval after rise and before fall) for all current sources is 1 psec. Both the rise time and fall time are 500 fsec. This trapezoidal pulse shape was chosen to approximate a Gaussian laser pulse. The maximal current supplied by the leftmost current source is 100 mA, the next source, on its right, supplies 70 mA peak current, and the rightmost source (in the last, $10^{th}$ photoelement) supplies 4.04 mA of peak current. All current sources are switched on and off simultaneously.

We have also done a simulation similar to the "F" case of Fig. 1, but collecting current from the left end of the transmission line, thus simulating pulse illumination from the semitransparent back contact, to which the dye sensitized nanocrystalline $TiO_2$ layer is attached, both physically and connected electrically, as an ohmic contact. We refer to this simulation as "back" (B) setup. The results of these two types of simulations are given in Fig. 2. The vertical axis is normalized to the maximum value of voltage at "Probe1" terminal (see Fig. 1) for both "F" and "B" simulations. It gives dimensionless values that are proportional to the current through the small resistor, R15, used as an ohmic load to the transmission line. The horizontal axis gives time in seconds, on a logarithmic scale.

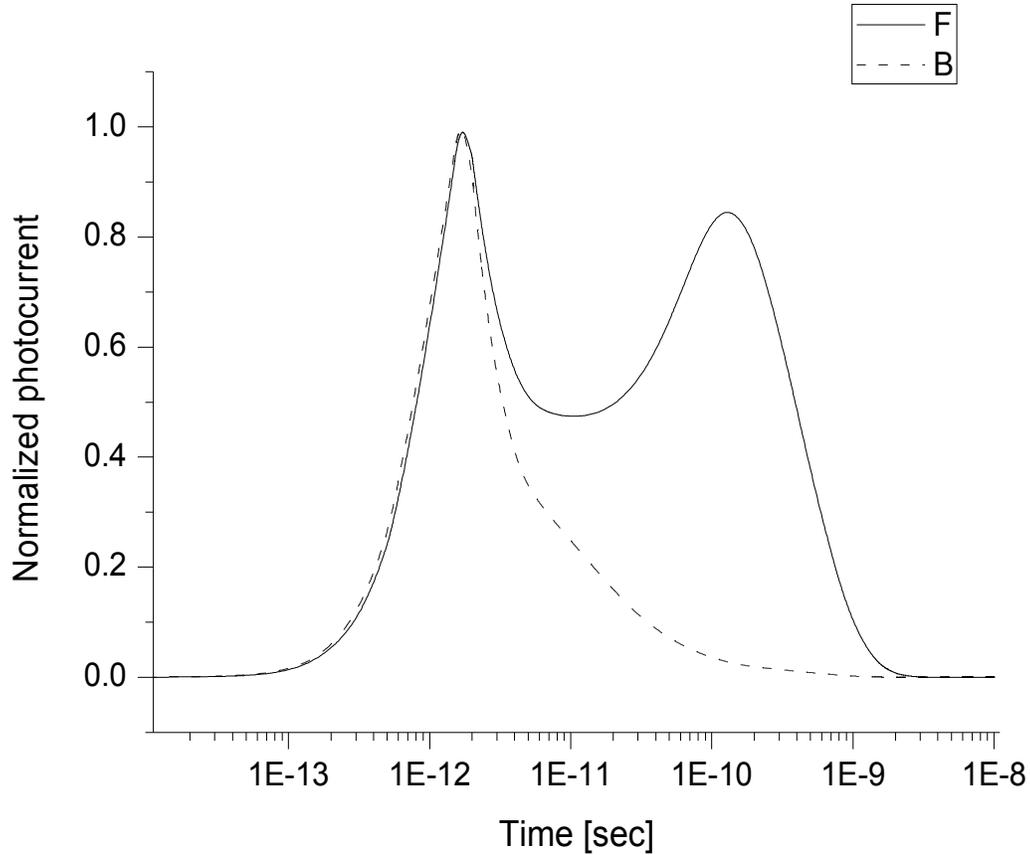

Fig. 2: Normalized (to the maximum value of each data set) simulated photocurrents from front pulse illumination (F) and back pulse illumination (B) for 10 current source units electrical transmission-line, given as a voltage drop on the small load resistor (R15 in Fig. 1).

The simulation result in Fig. 2 is very similar in its characteristics to both the experimental result and the results based on modeling diffusion and charge separation at the back contact, as reported by Schwarzburg et al [3]. In the front illumination case, there is a distinct, delayed second peak, that given the logarithmic time scale, represents the major part of the charge passed as the result of the illumination pulse. It is the result of larger photocurrents originated at the outer, intensely illuminated, layers of the dye sensitized polycrystalline semiconductor layer, before light intensity drops significantly upon reaching the back contact.

For thick dye sensitized polycrystalline semiconductor layers the light intensity reaching the back contact region can be expected to drop to very low levels, if the incident light intensity of front side illumination (the "F" case) is kept at the same value. In this case the first peak, representing the photocurrent response from the back contact region, can be expected to drop significantly. Simulations made with 20 current source units, instead of the 10 units electrical transmission-line in Fig. 1, show that the first peak is indeed suppressed relative to the second, delayed peak, as shown in Fig. 3. Compare this to the similar amplitudes of the "F" curve peaks in Fig. 2, for the "short" transmission line. The delay time is also increased, relative to the shorter transmission line, as expected.

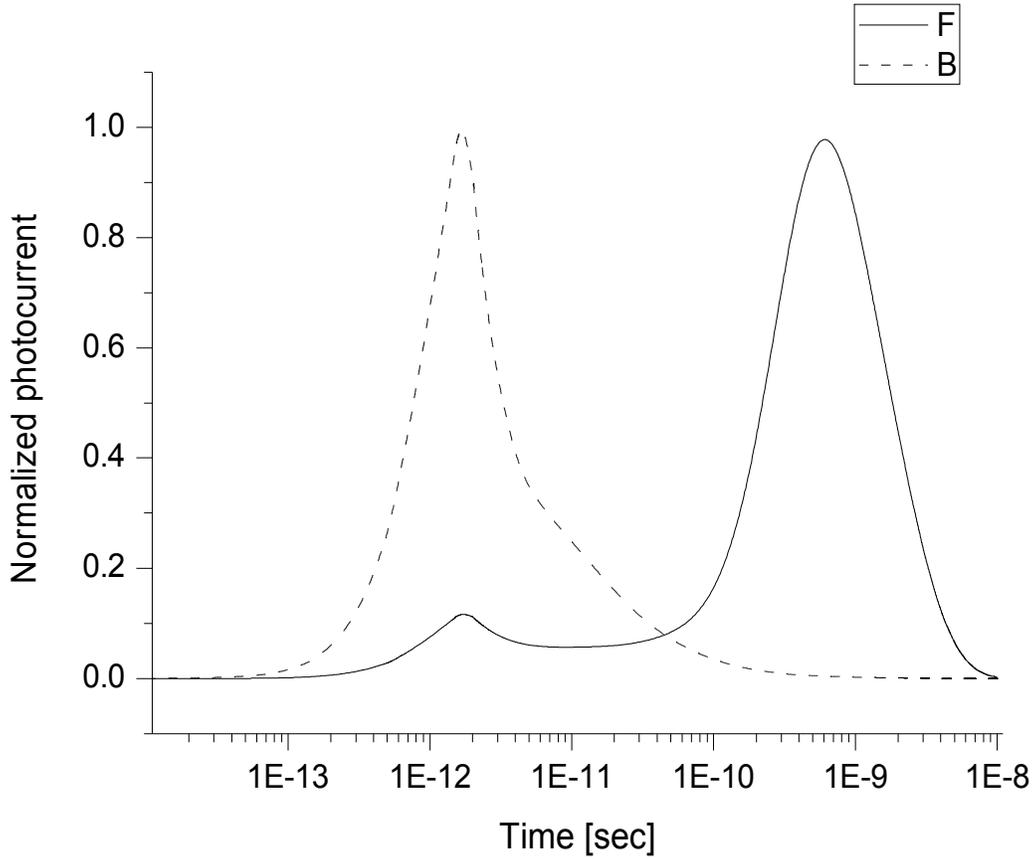

Fig. 3: Normalized (to the maximum value of each data set) simulated photocurrents from front pulse illumination (F) and back pulse illumination (B) of 20 current sources electrical transmission-line.

It should be noted that characteristic times (like the time separation between the two peaks in "F" curves) in the present simulation do not correspond to the actual kinetic parameters reported in ref [3] as no effort for parameter fitting has been made in choosing such combinations of RC time constants that would match the experimental results in ref. [3] quantitatively. We do, however get the qualitative characteristics of this system as exhibited by the two peak shape of the front illuminated photoresponse, and the single peak characteristics of the back illuminated setup. The very significant suppression of the first peak of the "F" curve relative to the second, delayed peak, as well as the increased separation in time between the two peaks, as the result of increasing the length of the transmission line, can suggest the use of experimentally obtained "F" and "B" curves for determining the thickness of the photo active layer. Differences in both the time separation between the two peaks of the "F" curve (about two orders of magnitude in Fig. 2, and almost three orders of magnitude in time in Fig. 3) and the ratio of peaks' intensity could be used by first constructing a calibration curve. Obtaining the "B" curve could help the positioning, on the time axis, of the location of the first peak, in particular in cases of very low intensity first peaks (thick layers).

Response characteristics, similar to those shown in Fig. 2, can be obtained using the circuit in the Fig. 4 below, where diode elements are substituted by elements consisting of a resistor and a capacitor connected in series. The small signal equivalent circuit in Fig. 4 had been tried after noticing that the transient voltages appearing on diodes' terminals of the circuit in Fig. 1 do not reach the values of forward bias at which diodes start conducting strongly in the forward direction.

Only when the maximal light intensity, as expressed by current pulse peak of the leftmost current source in both Fig. 1 and Fig. 4 circuits was set to 10 Amps (the large signal case), there was a clear difference between the transient responses of diodes' based circuit of Fig. 1, and its series-RC equivalent in Fig, 4. In the case of the large signal, the second (time delayed) peak in the front illuminated diode-based circuit has been suppressed to a relatively low, intensity plateau, while for the front illuminated series-RC equivalent circuit, the delayed peak was comparable, in amplitude to the first peak, similar to the small signal case for the response of the diode transmission line in Fig. 1, as shown in Fig. 2.

The suppression of the delayed peak in the case of diodes' circuit under high illumination intensity, (the large signal case) can be understood in terms of photocurrent's shunting by the leftmost (most intensely illuminated) diodes that had been switched on to the conducting state by being forward biased by transient voltages of ~1 V, as has been found during circuit simulations.

We are not aware of any experimental results showing the suppression of the delayed peak. As noted above, we postulate that peak suppression in the large signal case could occur by the mechanism of photocurrent shunting, or recombination, as the result of forward-biased diode switching on to the conductive state. If such-like peak suppression experimental results are found, they could provide an experimental support for the Helmholtz double layer model photogenerated charge separation suggested here.

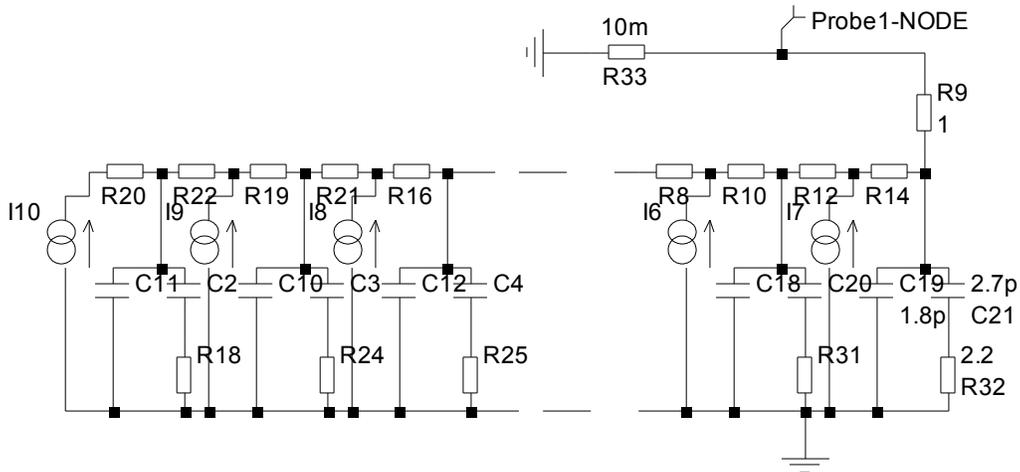

Fig. 4: The small signal, diodes' series-RC equivalent, of the circuit in Fig. 1, where all diodes are substituted by 2.2 ohm resistors and 2.7 pF capacitors connected in series.

**Discussion:**

The possibility of simulating the transient photoresponse of the dye sensitized nanocrystalline semiconductor photocell by use of photodiode elements arranged in a transmission line structure can suggest that, the taken for granted, fast injection of photogenerated electrons into the conduction band of nanocrystalline semiconductor ($TiO_2$ in our case) according to the, so called, "kinetic model" [5], can have its physical basis on fast kinetics provided by the electrochemical Helmholtz layer electric field.

The successful simulation of transient responses by the small signal equivalent of the circuit in Fig. 1, as given in Fig. 4, suggests the physical model of single trapping level along with diffusion as reported in [4]. However, both bulk traps and surface traps can be simulated by the small signal response of series-RC elements, the characteristic relaxation time constant of which, can simulate the life times of trapped charges. The Helmholtz double layer structure necessitates the presence of surface charges on semiconductor's surface, facing their counter charges in the electrolyte. Surface charges in surface traps could then be suitable, unshielded (from countercharges in the electrolyte) electrode-bound charge of the double layer. A conclusive support for the model of photogenerated charge separation by the electrical field of the surface double layer could be obtained from possible experimental evidence for the suppression of the delayed photoresponse peak at high illumination levels (the large signal case). Such a result could imply an existence of a rectifying junction at the semiconductor-electrolyte interface, as an electrochemical double layer analog of a MIM diode. If, on the other hand, the large signal experimental results do not show delayed peak suppression for the front illuminated case, the surface junction charge separation model would be irrelevant, leaving the transient response just characterizing the charge transport properties of the traps-modified diffusion model [4].